\documentclass[showpacs,preprintnumbers,amsmath,amssymb]{revtex4}

\usepackage{graphicx}
\usepackage{dcolumn}
\usepackage{bm}

\begin{document}

\preprint{APS/123-QED}

\title{Spin diffusion at finite electric and magnetic fields} 

\author{Y. Qi and S. Zhang}

\affiliation{Department of Physics and Astronomy, University of
Missouri-Columbia, Columbia, MO 65211} 

\date{\today}

\begin{abstract}
Spin transport properties at finite electric and magnetic fields are
studied by using the generalized semiclassical Boltzmann equation. 
It is found that the spin diffusion equation for non-equilibrium spin 
density and spin currents involves
a number of length scales that explicitly depend on the electric
and magnetic fields. The set of macroscopic equations can be used to 
address a broad range of the spin transport problems in magnetic multilayers
as well as in semiconductor heterostructure. A specific example of spin 
injection into semiconductors at arbitrary electric and magnetic fields 
is illustrated. 
\end{abstract}


\maketitle

Spin transport phenomena have been identified in various layered systems
including magnetic multilayers, magnetic tunnel junctions, and 
semiconductor heterostructure. The partial list of
these phenomena includes the spin injection-detection in metals
\cite{Johnson}, giant magnetoresistance \cite{Bass}, enhanced Gilbert damping
\cite{Bret,Bauer}, spin angular momentum
transfer \cite{Slon,Burhman} and spin current induced spin excitation
\cite{Tsoi}, ferromagnetic resonant \cite{Unknown}, tunnel magnetoresistance
\cite{Moodera}, and spin Hall effect in semiconductors \cite{Hirtsh,Zhang}.
Most recently, there are emerging interests in the spin injection and
detection in semiconductor spintronics \cite{Wolf}. Several theoretical
models based on a macroscopic drift-diffusion equation for each spin
channel has been proposed \cite{Flatte,Smith,Sarma}. 

The central observation of all these interesting spin-dependent transport
phenomena is the key role played by the electron spin. Unlike the physics
of strongly correlated systems, most of the spin transport phenomena 
can be understood in terms of the semiclassical description of electrons 
in the diffusive scattering limit as long as the spin degree of freedom
is properly being taken care of quantum mechanically. In this letter, we
construct a set of generalized spin dependent macroscopic equations from the
spinor form of the Boltzmann equation. We have found that these macroscopic
equations can self-consistently determine the charge density, spin density,
charge current and spin currents. The above mentioned spin dependent 
phenomena can be naturally understood by solving the macroscopic equations
with various experimental boundary conditions. In particular,
one can predict spin transport in semiconductor heterostructure
at arbitrarily strengths of electric and magnetic fields. 

Let us start with the spinor form of the semiclassical Boltzmann equation 
that describes the equation of motion for spin-half conduction electrons 
in an external electric field ${\bf E}$ and an effective magnetic field
${\bf H}_e $ 
\begin{equation}
\frac{\partial \hat{F}}{\partial t} + {\bf v} \cdot {\mathbf \nabla}_{\bf r}
\hat{F} + \frac{e}{m^*} ({\bf E} + {\bf v} \times {\bf H}_{e}) \cdot
{\mathbf \nabla}_{\bf v} \hat{F} - \frac{i}{\hbar} [ \hat{H} , \hat{F}]
= - \left(
\frac{\partial \hat{F}}{\partial t}
\right) _{colli}
\end{equation}
where $\hat{F}$ is the distribution function that depends on the position
${\bf r}$, momentum ${\bf p}$ (or velocity ${\bf v} = {\bf p}/m^* $), and
time $t$, ${\bf H}_{e} $ is the
effective magnetic field, $\hat{H}$ is the spin dependent Hamiltonian which
can be written in spin space as $\hat{H}= - \mu_B \mbox{\boldmath $\sigma
$ } \cdot {\bf H}_{e}$. The right-hand side of the equation is the 
collision term. 
Although the validity of the semiclassical Boltzmann equation is limited to 
the wave packet picture of the electron motion, the quantum mechanical nature
of the spin has been explicitly taken into account by the spinor form
of the distribution function and the spin dynamics is included in the 
commutator $[\hat{H} , \hat{F}]$ for the spin variable. 
 
Since we are interested in {\em spin-dependence} of the transport, we will
make a few simplifications to Eq.~(1). First, the effect of the magnetic field
on the orbital motion is discarded, i.e., we have set 
$ ({\bf v} \times {\bf H}_{e}) \cdot{\mathbf \nabla}_{\bf v} {\hat{F}}=0$; 
noticeably, this orbital term is in fact needed if one is interested
in the ferromagnetic resonance under an applied AC and DC magnetic fields
\cite{Unknown}. Second, 
we take the collision term in a simplest s-wave form by using 
momentum-independent relaxation time approximations for both non-spin-flip 
and spin-flip scattering, i.e.,
\begin{equation}
\left(
\frac{\partial \hat{F}}{\partial t}
\right) _{colli}
= \frac{\hat{F} - \bar{F}}{\tau} + \frac{2}{\tau_{sf}}
\left[
\bar{F} - \frac{\hat{I}}{2} {\rm Tr}\bar{F}
\right]
\end{equation}
where $\tau$ and $\tau_{sf}$ are the momentum and spin relaxation times,
$\bar{F} \equiv (1/4\pi) \int d\Omega_{\bf k} \hat{F} ({\bf r},{\bf v})$ is the
angular average over the ${\bf k}$-space, $\hat{I}$ is the 
$2\times 2$ unit matrix in spin space, and ${\rm Tr}$ represents the trace
over spin space. In ferromagnetic materials, the momentum relaxation time 
$\tau$ is also spin-dependent; this can be easily generalized by using a
$2\times 2$ matrix for $\tau$. For the notation simplicity, we write it
as a number. Finally, we should consider only the layered structure so that
the spatial dependence of the distribution is limited to one-dimension;
this is a practical approximation since the three-dimensional inhomogeneity
will lead to a complicated spatial-dependence of the current density that 
makes the calculation of the conductance extremely tedious. 
Let us denote the direction perpendicular
to the layered structure as $x$-direction. 

With above simplifications, Eq.~(1) becomes
\begin{equation}
\frac{\partial \hat{F}}{\partial t} + v_x \frac{\partial \hat{F}}{\partial x}
+ \frac{eE_x}{m^*} \frac{\partial \hat{F}}{\partial v_x} - \frac{\mu_B}{\hbar}
\mbox{\boldmath $\sigma$} \cdot ({\bf H}_e \times {\bf f} ) =
\frac{\hat{F} - \bar{F}}{\tau} + \frac{2 \bar{F} - \hat{I}  {\rm Tr}
\bar{F}}{\tau_{sf}}
\end{equation}
where we have introduced a spin-dependent distribution vector ${\bf f}$ by
separating the distribution function into spin dependent and 
spin independent parts, $\hat{F} \equiv
f_0 \hat{I} + \mbox{\boldmath $\sigma$} \cdot {\bf f}$, 
so that the commutator $(1/i\hbar) [ \hat{H}, \hat{F}]= 
({\mu_B}/{\hbar}) \mbox{\boldmath $ \sigma$} \cdot ({\bf H}_e \times {\bf f} )$.

Equation (3) is known as the generalized Bloch equation
\cite{Wilkins}. One of major tasks in
the spin transport theory is to solve this integro-differential equation
with desired precision for various physical structures. Here, we do not intend
to solve exactly the distribution function as a function of the momentum, 
position and time for a detail structure. Rather we construct 
a set of macroscopic equations from the above equation by integrating out
the momentum variable so that the spatial dependence of the macroscopic
spin dependent transport coefficients can be established for a wide range of
interesting physical systems.

The first set of the macroscopic equations are obtained by directly
summing over the momentum (velocity). By separately writing down
the spin dependent and spin independent parts of Eq.~(3), we have
\begin{equation}
\frac{\partial n_0 (x,t)}{\partial t} + \frac{\partial j_0 (x,t)}{\partial x} 
= 0
\end{equation}
and
\begin{equation}
\frac{\partial {\bf m} (x,t)}{\partial t} + \frac{\partial {\bf j}_m (x,t)}
{\partial x} - 
\frac{2 \mu_B}{\hbar} {\bf H}_{e}
\times {\bf m} (x,t) = - \frac{{\bf m} (x,t)}{\tau_{sf}}
\end{equation}
where we have defined the charge density $n_0 (x,t) = 
\sum_{\bf v} {\rm Tr}\hat{F}$,
the spin density ${\bf m} (x,t) = \sum_{\bf v}  {\rm Tr} (\mbox{\boldmath $
\sigma$ } \hat{F})$,
the charge current density $j_0 (x,t)
=  \sum_{\bf v} {\rm Tr} (v_x \hat{F})$, and
the spin current density ${\bf j}_m (x,t) = \sum_{\bf v}  {\rm Tr} (\mbox{
\boldmath $\sigma$} v_x \hat{F}) $. 
The above equations are exact macroscopic equations for charge and spin:
Eq.~(4) is the law of
charge conservation, and Eq.(5) governs the motion of spin in the presence of
spin relaxation due to spin flip scattering and spin precession
due to a magnetic field. 
To determine four macroscopic quantities ($n_0$, $j_0$,
${\bf m}$ and ${\bf j}_m$) we need to construct two more equations; this can
be furnished by timing $v_x$ at the both sides of Eq.~(3) and summing over
${\bf v}$. Again, after separating the spin dependent and spin independent
parts of Eq.~(3), we find
\begin{equation}
\frac{\partial j_0 (x,t)}{\partial t} + \overline{v_x^2} 
\frac{\partial n_0 (x,t)}{\partial x} + \frac{e E_x}{m^*} n_0 (x,t)
= - \frac{j_0 (x,t)} {\tau}
\end{equation}
and 
\begin{equation}
\frac{\partial {\bf j}_m (x,t)}{\partial t} + \overline{v_x^2}
\frac{\partial {\bf m} (x,t) }{\partial x} + \frac{e E_x}{m^*} {\bf m}
(x,t) - \frac{2 \mu_B}{\hbar} {\bf H}_e \times {\bf j}_m (x,t)
= - \frac{{\bf j}_m (x,t) }{\tau}. 
\end{equation}
In deriving the above closed form, we 
have used the ``mean field'' approximation $\sum_{\bf v} v_x^2 \frac{\partial 
\hat{F}}{\partial x} \approx  \overline{v_x^2} \sum_{\bf v}  \frac{\partial
\hat{F}}{\partial x}$. In another words, we only consider the first
two polynomials (zeroth and first orders) of the distribution function
in the expansion respect to 
the direction of the momentum. This approximation has 
been used for most of diffusion theories and it is valid as long as the
system is not too much anisotropic. In principle, however, we should keep
higher orders of polynomials and one requires to use more macroscopic
variables such as ${\bf Q}_n = \sum_{\bf v} v_x^n {\rm Tr} 
( \mbox{\boldmath $\sigma$ } \hat{F} )$
where $n=0$ is for the spin (number) density, 
$n=1$ is for the spin current density, $n=2$ is for the spin energy density, 
and so forth. Truncating the expansion at $n=2$ makes the equations
for the macroscopic variables in the closed form, but it
sacrifices the accuracy in the solution of the Bloch equation for anisotropic
systems. However, we argue that inclusion of higher orders does not in fact
improve much of the accuracy of the
calculation since we have already made an approximation in deriving
the Bloch equation by assuming a ${\bf k}$ independent relaxation time.
  
Equations (4) and (6) describe the non-equilibrium
charge density and charge current density. Alone with the Poisson
equation, Eqs.(4) and (6) completely determine the charge carrier distribution
in a layered structure in the presence of arbitrary electric fields. 
A well-known example is the semiconductor p-n junction
where the electric field in the depletion region
is quite large, thus the charge diffusion 
should be calculated by keeping all orders of the electric field
in solving Eqs.~(4) and (6).
Here our focus is on the spin part of the macroscopic
equations, Eqs.~(5) and (7).

The spin diffusion equation 
can be readily derived by expressing the spin current density
in terms of the spin density from Eq.~(7) and by inserting the
resulting expression into Eq.~(5). Let us first consider the time 
independent case. As Eq.~(7) contains a cross product of the spin current
density vector and the magnetic field, it
is convenient to separate the components parallel and perpendicular
to the magnetic field, i.e.,
\begin{equation}
j_m^z = - \tau \overline{v_x^2} \frac{\partial m_z}{\partial x}
- \frac{e E_x \tau}{m^*} m_z
\end{equation}
and 
\begin{equation}
j_m^{\pm} = - \frac{1}{1 \mp 2i \hbar^{-1} \mu_B\tau H_{e}} \left(
\tau \overline{v_x^2} \frac{\partial m_{\pm}}{\partial x}
+ \frac{e E_x \tau}{m^*} m_{\pm}  \right)
\end{equation}
where we have assumed a uniform magnetic field in the z-direction, and defined
the perpendicular components of the spin current density 
$j_m^{\pm} = j_m^x \pm i j_m^y$ and the spin density
$m_{\pm} = m_x \pm i m_y$. By placing them into Eq.~(5), we arrive
at the spin diffusion equations for the longitudinal and transverse spin
density, 
\begin{equation}
\frac{\partial^2 m_z}{\partial x^2} + \frac{{\rm Sign}(E_x)}{\lambda_E} 
\frac{\partial m_z}{\partial x} + \left(
\frac{1}{\lambda_E \lambda_{c}} - \frac{1}{\lambda_{sdf}^2} 
\right) m_z = 0
\end{equation} 
and 
\begin{equation}
\frac{\partial^2 m_{\pm}}{\partial x^2} +  \frac{{\rm Sign}(E_x) }
{\lambda_E} \frac{\partial
m_{\pm}}{\partial x} + \left( \frac{1}{\lambda_E \lambda_{c}}
+ \frac{1}{\lambda_h^2} - \frac{1}{\lambda_{sdf}^2} \mp \frac{i}{\lambda_s^2}
\right)
m_{\pm} = 0
\end{equation}
where ${\rm Sign}(E_x) = 1 (-1) $ for $E_x > (<) 0 $. We have introduced 
a number of relevant lengths: the usual spin-flip scattering length 
$\lambda_{sdf} = \sqrt{\overline{v_x^2} \tau \tau_{sf}}$, 
the length associated with acceleration of electrons by
the electric field $\lambda_E = m^* \overline{v_x^2}/e|E_x|$, the length
due to non-uniform electric field (or due to charge accumulation in the
depletion layers) $\lambda_c^{-1} = |(1/E_x) dE_x/dx| $, 
the spin precession length induced by the magnetic field $\lambda_h =
\hbar \sqrt{\overline{v_x^2}}/2\mu_B H_e $, and the precessional spin density 
decay length $\lambda_s = \sqrt{\hbar \tau \overline{v_x^2}/2\mu_B H_e} .$
  
The spin diffusive equations, Eq.~(10) for the longitudinal spin density
and Eq.~(11) for the transverse spin density, are our central results.
For the longitudinal spin density, the spin diffusion is not affected
by the magnetic field since we have disregarded the electron orbit motion.
The electric field as well as its spatial derivative have significant effects 
on the spin diffusion for 
semiconductor heterostructure where the interface regions are expected
to possess large space charges and thus large electric fields. For example,
for a non-degenerate semiconductor, $m^* \overline{v_x^2}
\approx k_B T$, and thus
$\lambda_E $ is only about 100 \AA $\,$ for a typical electric field of
$2.5 \times 10^4$ (V/cm) at T=300K. In a moderated doped interface, the gradient
of the electric field in the depletion layer is large so that 
$\lambda_c$ is also of the order of 100 \AA . Such small $\lambda_E$ and
$\lambda_c$ make the diffusion length
highly anisotropic: the spin diffusion is much easier when the direction
of the diffusion is same as the direction of the electric field; otherwise,
the spin diffusion is prohibited. For a metallic layer, however, the
electric field effect is small since $\lambda_E$ is typically of the
order of $10 \mu m $ if we take $m^* \overline{v_x^2} \approx 1 $ (eV) 
(a fraction of the Fermi energy) and $E_x$ is of the order of $10^3$ (V/cm)
for the extremely high current density. Thus, the
commonly used spin diffusion equation
which neglects the electric field is justified
for metallic multilayers.

We point out that a similar spin diffusion equation for the
longitudinal spin density has recently derived from the spin-dependent
drift-diffusion equations by Yu and Flatt\'{e} \cite{Flatte},
and Albrecht and Smith \cite{Smith}
with no magnetic fields. Our derivation from the
Boltzmann equation generalizes their results significantly.
First, we are able to address the longitudinal and transverse
spin diffusion at any magnetic field; the spintronics application is
inevitably involving the magnetic field. Second, we include the
additional length scale $\lambda_c$ due to non-uniformity of electric fields
at the depletion layers; this effect has been neglected all together in the
previous treatment. Third, we have started from the semiclassical 
Boltzmann equation so that
we have a controlled access on the approximations made in deriving the
spin diffusion equations; the deviation of the macroscopic equations
from the exact solutions of the Boltzmann equation
can be compared in various cases \cite{thesis}.

We now turn to a discussion on the transverse spin density and spin 
current density.
The spin diffusion equations involve two more length scales associated with
the magnetic field in addition to the length scales in the longitudinal
diffusion equation. The total spin current density decaying length 
would depend on the detail magnitudes of these lengths scales. 
The magnetic field enters
in two places: one is to compete with the spin flip scattering (notice
the positive sign of $1/\lambda_h^2$) and the other is a precessional
decay of spin density due to scattering and precession (notice the imaginary
$i$). To give out a 
quick estimation of the magnetic field on the transverse spin diffusion, 
we take an example of the non-degenerate semiconductor again.  
We find that the magnetic field of the order of 1000 (Oe) can make
$\lambda_h$ and $\lambda_s$ comparable to $\lambda_{sdf}$. Therefore,
one must consider the spin diffusion by including both finite electric
and magnetic fields in a realistic spintronics device. 

Our generalized spin diffusion equations
in the presence of finite electric and magnetic fields 
can be used to a broad range of spin transport phenomena. By taking the
limiting cases, many previously known spin diffusion phenomena are
recovered. The giant magnetoresistance for currents perpendicular to the 
plane of the layers is a simplest example where both electric and
magnetic fields are assumed small \cite{Johnson}. The phenomenon of
the spin angular momentum transfer with non-collinear 
magnetization vectors of two ferromagnetic layers 
can be studied by taking the limit of a large internal 
magnetic field and a vanishingly small electric field \cite{Zhang2}.
The ferromagnetic resonant experiments can be analyzed via the
reduction of the diffusion constant due to a large magnetic 
field \cite{Unknown}.
One can also derive a solution of the time-dependence of the spin accumulation  
in magnetic multilayers by keeping the time dependence diffusion term in
Eq.~(11). 

Finally, we illustrate an important application of our spin diffusion equation
in a ferromagnetic/tunnel barrier/semiconductor heterostructure. To make our
calculation much simpler, we assume a uniform electric field near
the interface of the insulator barrier and the semiconductor; a detailed
calculation on the self-consistently determined non-uniform electric fields
due to depletion potentials and the external bias will be published
elsewhere. In the present case, the solutions of the longitudinal and the
transverse spin density, Eqs.~(10) and (11), are simple 
exponential functions with the exponential decaying factors given by
the above mentioned length scales. To determine absolute values of the
spin density and spin polarization of the current, we need to use boundary
conditions across the insulator barrier. Let us introduce a spin-dependent
tunnel resistance across the barrier as  
$\hat{R} = R_0 (\hat{I} +P \mbox{\boldmath $\sigma$} \cdot
{\bf M}_F)$ where ${\bf M}_F$ is the unit vector in
the direction of the ferromagnetic
magnetization and $P$ is the tunneling polarization. The conditions for
continuous of spin density and spin current density at the interface between
the semiconductor and the insulator barrier completely determine the
prefactors of the exponential solutions.

\begin{figure}[h]
\begin{center}
\includegraphics[width=3.5in]{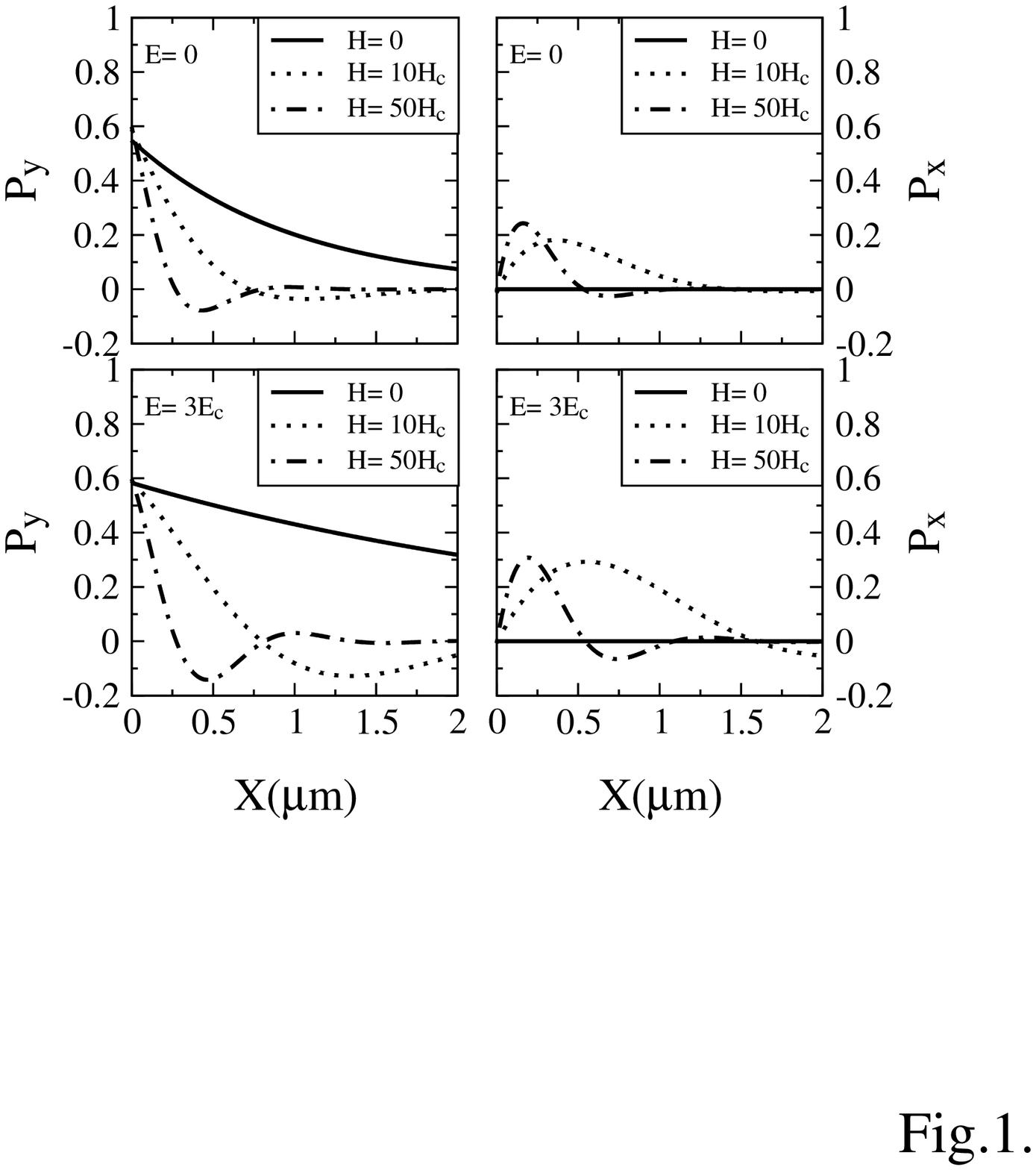}
\renewcommand{\figurename}{Fig.}\caption {
Spin-polarization of the current as a function of position
in the semiconductor for different electric and magnetic fields. The
magnetic field is applied at z-axis and the ferromagnet is polarized
at $y$ direction for all panels in the figure.
$P_x \equiv j_{\bf m}^x/j_0 $ and $P_y \equiv j_{\bf m}^y/j_0 $ are $x$ and
$y$ components of the spin polarization of the current, respectively.
The parameters are: the spin-diffusion length
of the semiconductor is $\lambda_{sdf}= 1.0 \mu m$, the critical electric field
and magnetic field are defined as $E_c \equiv k_B T/e\lambda_{sdf}$
and $H_c = \hbar /2 \mu_B \tau_{sd}$, the resistance
ratio $\rho_{semi} \lambda_{sdf} /R_0 = 0.1$ (where $\rho_{semi}$ is the
resistivity of the semiconductor), and the polarization of the
barrier resistance is $P=0.6$.
}
\end{center}
\end{figure}

In Fig.~1, we show the non-equilibrium spin polarization of the current density
as a function of the position in the semiconductor in different electric
field and magnetic fields. Here we assume ${\bf M}_F$ is at the $y$ direction  
and the magnetic field is applied at the z-direction. At zero magnetic field,
the electric field can enhance or reduce the spin density, depending on 
the direction of the field, as pointed out in \cite{Flatte, Smith}. 
The magnetic field has dramatic effects on
the spin density and the spin polarization; the spin diffusion length
is shorten significantly even for a moderate magnetic field, say, 1000 (Oe).
On the other hand, one also develops a spin current
in $x$-direction which
was absent without the magnetic field. This $x$-component of the spin
density and spin current comes from the rotation of the injected spin
(originally at $y$ direction) by the magnetic field.

In summary, we have presented a set of macroscopic equations for
charge density, spin density, charge current and spin current at finite
electric and magnetic fields. These equations are the generalization of 
previous useful spin diffusions at various limiting cases. These equations
can be applied to a wide range of the diffusive spin transport phenomena.
This work is supported by National Science Foundation (DMR-0076171).

%

\end{document}